\documentclass{PoS}

\newcommand{\sax}{{\textit{Beppo\-SAX}}}
\newcommand{\gro}{{\textit{CGRO}}}
\title{X-ray spectral states of microquasars}

\ShortTitle{X-ray spectral states of microquasars}

\author{\speaker{Julien Malzac} and Renaud  Belmont\\
        CESR, Toulouse, France\\
        E-mail:  \email{Julien.Malzac@cesr.fr},
        \email{Renaud.Belmont@cesr.fr}}

\abstract{

We discuss the origin of the dramatically different X-ray spectral shapes observed in the Low Hard State (LHS: dominated by thermal comptonisation) and  the High Soft State (HSS: dominated by the accretion disc thermal emission and non-thermal comptonisation in the corona). We  present numerical simulations using a new code accounting for the so-called synchrotron boiler effect.  These numerical simulations when compared to the data  allow us to constrain the magnetic field and temperature of the hot protons in the corona. 
For the hard state of Cygnus X-1 we find a magnetic field below equipartition with radiation,  suggesting  that the corona is not powered through magnetic field dissipation (as assumed in most accretion disc corona models). On the other hand, our results also point toward proton temperatures that are substantially lower than typical temperatures of the ADAF models. Finally, we show that in both spectral states  Comptonising plasma could be powered essentially through power-law acceleration of non-thermal electrons, which are then partly thermalised by the synchrotron and Coulomb boiler. This suggests that, contrary to current beliefs, the corona of the HSS and that of the LHS could be of very similar nature. The differences between the LHS and HSS coronal spectra would then be predominantly caused by the strong disc soft cooling emission which is present in the HSS and absent in the LHS. }

\FullConference{VII Microquasar Workshop: Microquasars and Beyond\\
		 September 1-5 2008\\
		 Foca, Izmir, Turkey}

\begin{document}
\section{Introduction}\label{sec:intro}

Black hole binaries are observed in two main spectral states. At luminosities exceeding a few percent of the Eddington luminosity\footnote{For hydrogen gas, the Eddington luminosity is $L_{\rm E}= 1.3 \times 10^{38} M$ erg s$^{-1}$ where $M$ is the black hole mass expressed in units of solar masses} ($L_{\rm E}$), the spectrum is generally dominated by a thermal component peaking at a few keV which is believed to be the signature of a geometrically thin optically thick accretion disc (Shakura \& Sunyaev 1973).
At higher energies the spectrum is non-thermal and usually presents a weak steep power-law component (photon index $\Gamma \sim 2.3-3.5$) extending at least to MeV energies, without any hint for a high energy cut-off. This soft power law is generally interpreted as inverse Compton up-scattering of soft photons (UV, soft X)  by an hybrid thermal/non-thermal distribution of electrons in a hot relativistic plasma (the so-called ``corona").  The electron distribution is then characterised by a temperature $\sim$ 30--50 keV, a Thomson optical depth $\sim$ 0.1-0.3 and a quasi power-law tail ($n(\gamma)\propto \gamma^{-s}$, with index $s$  in the range 3.5--4.5) which is responsible for most of the hard X-ray and $\gamma$-ray  emission (Gierli{\'n}ski et al. 1999, hereafter G99; Frontera et al. 2001a; McConnell et al. 2002, hereafter MC02; Del Santo et al. 2008).
Since in this state the source is bright in soft X-rays and soft in hard X-rays it is called the High Soft State (hereafter HSS).

At lower luminosities (L$<$ 0.01 $L_{\rm E}$), the appearance of the accretion flow is very different: the spectrum can be modelled as a hard power-law $\Gamma \sim 1.5-1.9$
  with a cut-off at $\sim100$ keV. The $\nu F_{\nu}$ spectrum then peaks at around a hundred keV. 
 Since  the soft X-ray luminosity is faint and the spectrum is hard, this state is called the Low Hard State (hereafter LHS). LHS spectra are generally very well fitted by  thermal Comptonisation, i.e. multiple Compton up-scattering of soft  photons by a Maxwellian distribution of  electrons (see Sunyaev \& Titarchuk 1980) in a hot ($kT_{\rm e}\sim$ 50--100 keV)  plasma of Thomson optical depth $\tau_{\rm T}$ of order unity. 
Although a weak non-thermal component is clearly detected at MeV energies in the LHS of Cygnus X-1 (MC02) and GX~339--4 (Wardzi{\'n}ski et al. 2002; Joinet et al. 2007) most of the hard X-ray luminosity emerges in the form of Comptonisation by a thermal electron distribution.  Spectral fits with hybrid thermal/non-thermal Comptonisation models suggest the slope of the non-thermal tail  in the electron distribution is steeper ($s>5$)  than in the HSS (see MC02). 

In brief, both spectral states are well represented by Comptonisation by an hybrid electron distribution. In the LHS the temperature and optical depth of the thermal electrons are higher, and the slope of the non-thermal tail seem steeper than in the HSS.  Consequently, the X-ray emission is dominated by thermal Comptonisation in the LHS and by non-thermal Comptonisation in the HSS.

Assuming  that this interpretation of the spectra is correct, we present a relatively simple coupled kinetic-radiation model that  allows us to understand the origin of the very different spectral shapes observed in the two spectral state as well as  the spectral evolution during state transitions (see e.g. Del Santo et al. 2008). This model is independent of any dynamical accretion flow model or geometry but when compared to the data it can be used to constrain the physical parameters of the emitting corona such as the magnetic field or the temperature of hot protons. In this paper, we summarise our main results while a thorough investigation of the model and its discussion in the context of the observations will be presented in Malzac \& Belmont (2008).
 
\section{Model}\label{sec:model}

We use the code of Belmont et al. (2008 see also Belmont et al. in  these proceedings). This code solves the kinetic equations for photons, electrons and positrons  accounting for Compton scattering (using the full Klein-Nishima cross section), electron-positron pair production and annihilation, Coulomb interactions (electron-electron and electron-proton), synchrotron emission and absorption and $e$-$p$ bremsstrahlung. Radiative transfer is dealt using a usual escape probability formalism. 

We model the Comptonising region has a sphere with radius $R$  of fully ionised proton-electron magnetised plasma in steady state. The Thomson optical depth of the sphere is $\tau_{\rm T}=\tau_{\rm i}+\tau_{\rm s}$, where $\tau_{\rm i}=n_i\sigma_{\rm T}R$ is the optical depth of ionisation electrons (associated with protons of density $n_i$) and $\tau_s=2n_{\rm e^{+}}\sigma_{\rm T}R$ is the optical depth of electrons and positrons due to pair production ($n_{\rm e^{+}}$ is the positron number density). $\sigma_{\rm T}$ is the Thomson cross section.
 The radiated power is quantified through the usual compactness parameter  $l=\frac{L\sigma_{\rm T}}{R m_{\rm e} c^3}$,
where $L$ is the luminosity of the Comptonising cloud,  $m_{\rm e}$ the electron rest mass and $c$ the speed of light. 

We will consider three possible channels for the energy injection in our coupled electron-photon system: 
\begin{enumerate}
\item Non-thermal electron acceleration  with a compactness $l_{\rm nth}$.
We model the acceleration process by assuming electrons are continuously injected with a power-law distribution of index $\Gamma_{\rm inj}$, between Lorentz factor $\gamma_{\rm min}$ and $\gamma_{\rm max}$ (i.e. $n(\gamma)\propto \gamma^{-\Gamma_{\rm inj}}$). Observations of X-ray binaries in the soft states indicate $\Gamma_{\rm inj}$  in the range 2.0--3.5 (G99). 

\item  Coulomb heating with a compactness $l_{\rm c}$. 
Electrons are supposed to interact by Coulomb collisions with a distribution of thermal protons. When the protons have a larger temperature than the electrons, the electrons gain energy and $l_{\rm c}>0$ (see below Eq. \ref{eq:coul}). In our model, we fix $l_{\rm c}$,  then determine the proton temperature corresponding to this compactness and to the exact steady state electron distribution.

\item External soft radiation coming from the geometrically thin accretion disc and entering the corona with a compactness $l_{\rm s}$, that we model as homogeneous injection of photons with a blackbody spectrum of temperature $kT_{\rm bb}$.
\end {enumerate}
Since, in this model, all the injected power ends up into radiation we have: $l=l_{\rm nth}+l_{\rm c}+l_{\rm s}$ in steady state. In addtition the magnetic field $B$ is parametrized through the usual magnetic compactness: $l_{B}=\frac{\sigma_{\rm T}}{m_{\rm e} c^2}R \frac{B^2}{8\pi}$.

\begin{figure*}
 \includegraphics[width=\textwidth]{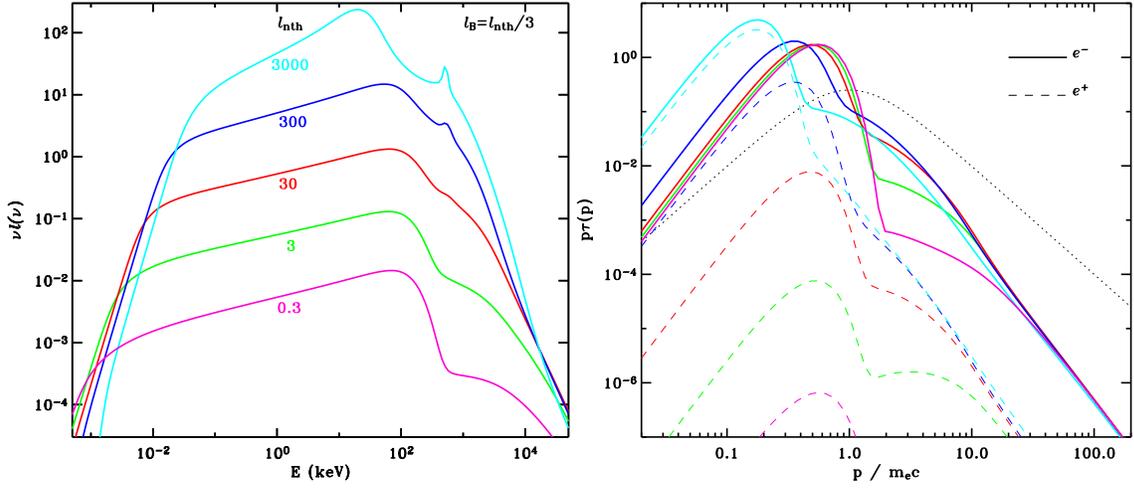}
 \caption{Synchrotron self-Compton models with pure non-thermal injection. Effects of varying $l_{\rm nth}$ at constant $l_{\rm nth}/l_{B}$. Simulated photon spectra (left panel) and particle distributions for $l_{\rm nth}$= 0.3 (pink), 3 (green), 30 (red),  300 (blue),  3000 (cyan) and $l_{\rm nth}/l_{B}=3$. In the right hand side panel the full curves show the electron while the dashed ones show the positron distributions. { The dotted curve shows the shape of the injected electron distribution.}}
 \label{fig:le}
\end{figure*}

\section{Simulation Results}\label{sec:simres}

\subsection{Synchrotron self-Compton models with pure non-thermal injection}\label{sec:nonthe}

In this section we assume that the protons are cold ($l_{\rm c}=0$) and the external photons are neglected ($l_{\rm s}=0$).
Figure~\ref{fig:le} shows the dependence of the photon spectrum on $l_{\rm nth}$ for  $l_B=l_{\rm nth}/3$, which corresponds to approximate equipartition  of magnetic field with radiation.  
The other fixed parameters are $\Gamma_{\rm inj}=3$, $R=5 \times 10^7$ cm,  $\tau_{\rm i}=2$, $\gamma_{\rm min}=1$ and $\gamma_{\rm max}=1000$.
For a wide range of compactness the spectrum peaks around 65 keV and and the X-photon index is $\Gamma\simeq1.7$.  { The right panel of Fig~\ref{fig:le} shows the steady state particle distribution corresponding to these photon spectra. These particle distributions are all  thermal at low energies with a non-thermal tail at high energies.  
The non-thermal electrons produce self-absorbed synchrotron radiation peaking around a few eV, at the synchrotron turn-over frequency.  The hybrid thermal non-thermal distribution then Compton up-scatters the self-absorbed synchrotron radiation, forming spectra extending into the hard X-ray and $\gamma$-ray domain.  These spectra are dominated by the Comptonisation by the thermalised particles. The peak and energy of the cut-off depends on the temperature.} In the following we will refer to this process as Synchrotron Self-Compton emission (or SSC). 

In terms of particle kinetics, this can be understood as follows. The injected power-law electrons cool down by emitting { SSC radiation and by interacting through Coulomb collisions with lower energy particles.  Below a critical Lorentz factor $\gamma_{\rm t}$, the particles emit synchrotron in the self-absorbed regime: their emission is immediately absorbed by lower energy electrons. This very fast process of synchrotron emission and self absorption represents a very efficient and fast thermalising mechanism. Very quickly the low energy part of the electron distribution forms a quasi Maxwellian. This synchrotron boiler effect is able to transfer  a large fraction of the non-thermal energy of the electrons (of Lorentz factor $<\gamma_t$) into heating of the lowest energy particles, keeping the effective temperature around 40 keV. This is the so-called synchrotron boiler effect (Ghisellini, Guilbert and Svensson 1988).

The Lorentz factor $\gamma_t$ corresponds to the energy of the electrons radiating most of their synchrotron luminosity around the synchrotron turn over frequency  $\gamma_t\simeq10$, the exact value being rather insensitive to the model parameters. Not only the electrons at or above  $\gamma_{\rm t}$ do not contribute to the heating of the lower energy particles but their soft synchrotron radiation also provides the dominant contribution to the Compton cooling of the plasma. Therefore the efficiency of the synchrotron boiler as a heating mechanism will depend strongly on the relative fraction of electrons present  above and below $\gamma_{\rm t}$.}  { At the same time Coulomb collisions between high energy and low energy electrons also plays important (and possibly dominant)  thermalising role, similar to that of the synchrotron boiler.}

{ As the ratio $l_{B}/l_{\rm nth}$ is fixed, both particle heating and radiative cooling scale with $l$ and the spectra depend only weakly on compactness. At low compactness the small differences between the  spectra of various compactness are entirely due to the increasing importance of Coulomb effects at low $l_B$.
At very large $l_{\rm nth}$, pair production effects become important and electron positron pairs contribute significantly to the total optical depth ($\tau_{\rm T}\simeq9$ for $l_{\rm nth}=3 \times 10^3$). As a consequence the effective temperature decreases, as more particle share the boiler energy, but the Compton parameter actually increases producing even harder spectra ($\Gamma$=1.39 for  $l_{\rm nth}=3 \times 10^3$). We also note that, at high compactness, pair annihilation leads to the formation of the line apparent around 511 keV  in some spectra of Fig.~\ref{fig:le}.

These simulations show that  even though the electron injection is purely non-thermal, most of the radiated luminosity emerges as quasi-thermal Comptonisation. The basic spectral properties of black hole binaries in the LHS appear to be produced over a wide range of luminosity (at least up to 0.3$ L_{\rm E}$) through pure non-thermal electron injection in a magnetised plasma, with no need for additional heating mechanism nor soft photons.

\subsection{Effects of external soft photons}\label{sec:ls}

In addition to soft photons produced internally through synchrotron, the hot comptonising plasma may intercept a fraction of  the thermal radiation from the accretion disc.  We denote as $l_{\rm s}$ the compactness associated to this additional soft photon injection. The main effect of the external soft photons field, is to increase the Compton cooling rate  of the leptons. The equilibrium temperature of the thermal component is lower than in pure SSC models. If $l_{\rm s}$ is strong ($l_{\rm s}\sim l_{\rm nth}$) the thermalised leptons are so cool that most of the luminosity is radiated by the non-thermal particles. Basically the effect of $l_{\rm s}$, is to increase the fraction of radiation emitted through non-thermal Comptonisation.    
  
 \begin{figure*}
 \includegraphics[width=\textwidth]{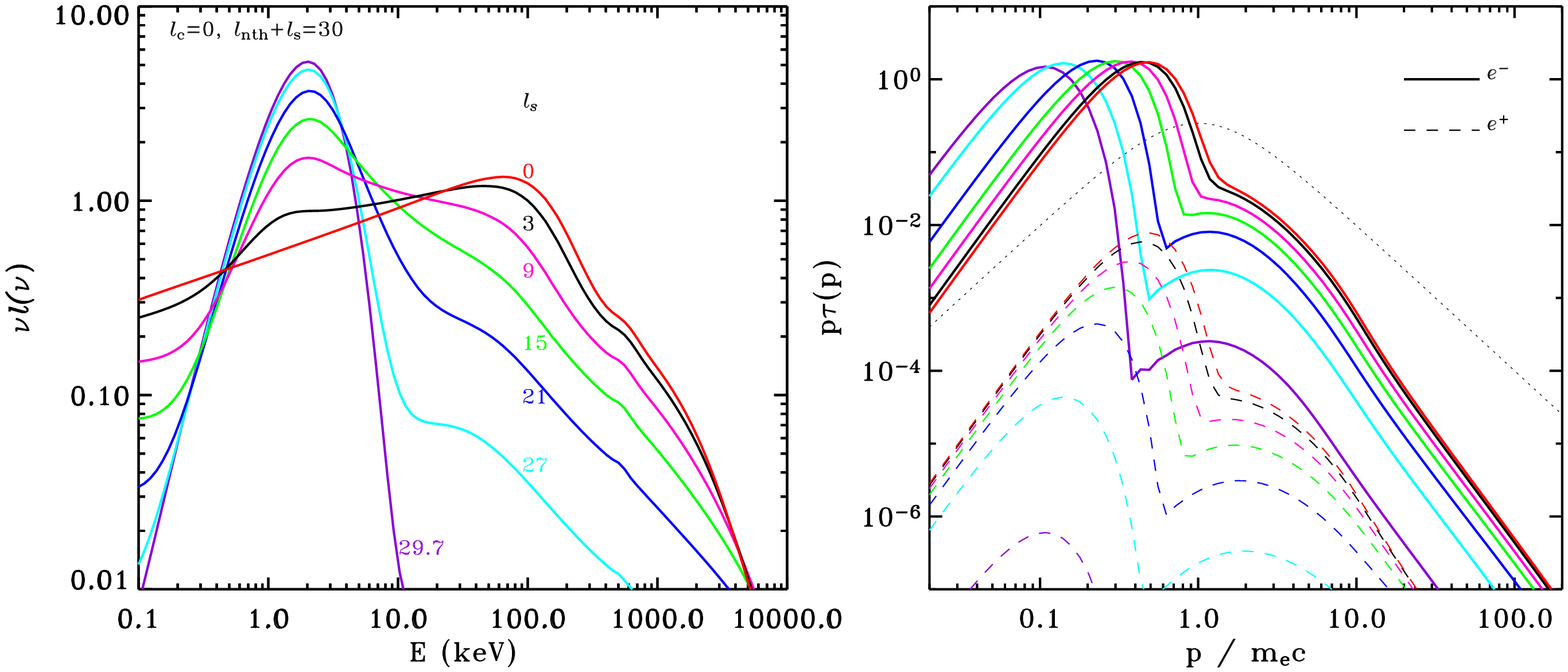}
 \caption{{ Pure non-thermal models with external soft photons.} Effects of varying the soft photon compactness $l_{\rm s}$ at constant $l$ on the photon spectrum  (left hand side) and the particle distributions (right hand side).  The red curve shows our fiducial model obtained for $l_{\rm s}=0$, $l_{\rm nth}=30$, $l_B=10$, $\Gamma_{\rm inj}=3$, $\gamma_{\rm min}=1$, $\gamma_{\rm max}=1000$, $\tau_{\rm i}=2$, $R=5 \times 10^7$ cm. The other curves show a sequence of models obtained  keeping $l=l_{\rm s}+l_{\rm nth}=30$ while the soft photon compactness $l_{\rm s}$ is varied.  The black, pink, green, blue, cyan and purple curves show the results for $l_{\rm s}$=3, 9, 15 21, 27, 29.7 respectively.  The soft photons are injected with a blackbody spectrum of temperature $kT_{\rm bb}$=213.5 $l_{\rm s}^{1/4}$ eV. In the right hand side panel, the full curves show the electron while the dashed ones show the positron distributions.}
 \label{fig:lsoft}
\end{figure*}

Fig~\ref{fig:lsoft} shows a sequence of spectra, starting from our fiducial case $l_{\rm nth}=30$, $l_{B}=10$ in which  $l_{\rm s}$  is increased keeping the total compactness $l=l_{\rm s}+l_{\rm nth}$ constant. As the soft photons flux rises in the corona, inverse Compton increasingly dominates the cooling of the non-thermal particles eventually turning the synchrotron boiler off. As $l_{\rm s}$ is increased,  the electron distribution cools down to the Compton temperature. The emissivity of the thermal component is strongly reduced and the hard X-ray spectrum becomes gradually dominated by the emission of the cooling non-thermal distribution. Then, the non-thermal emission vanishes as $l_{\rm nth}$ decreases.

This spectral sequence, where $l_{\rm s}$ is increased,
is reminiscent of that observed during the state transition of accreting black holes (DGK07; Del Santo et al. 2008).  This indicates that cooling by the soft photons from the accretion disc probably plays a major role in the changing appearance of the corona during such events.  It was known from spectral fits with the hybrid thermal-non-thermal Comptonisation model {\sc eqpair} (Coppi 1992) that due to the strong thermal emission from the disc  that the ratio of coronal heating compactness over soft photon compactness $l_{\rm h}/l_{\rm s}$ was lower in the HSS, implying a lower temperature of the plasma. Here we show that the higher $l_{\rm nth}/l_{\rm h}$ ratio of  the HSS is naturally explained with this synchrotron boiler model. 
However as will be discussed in Section~\ref{sec:cygx},  real states transitions in X-ray binaries are a bit more complex and the spectral changes cannot be entirely attributed to a change in the soft photon compactness.

\subsection{Effects of $e$--$p$ Coulomb heating}\label{sec:effectsofCoulombheating}

In this section we investigate the effects of Coulomb heating by a thermal distribution of protons in a two temperature plasma. Fig~\ref{fig:lcoul} shows the evolution of the spectrum when $l_{\rm c}$ is varied  at fixed $l=l_{\rm nth}+l_{\rm c}=30$,  $l_B=l/3$, $l_{\rm s}=0$. The other parameters are those of our fiducial model. When the fraction of Coulomb heating increases the effects of the non-thermal component both in terms of high energy emission and soft cooling photon flux are reduced, as a consequence  the temperature of the thermal component increases. This leads to an increasingly  hard  spectrum peaking at increasingly large photon energies.  Finally, when Coulomb heating dominates the equilibrium electron distribution is very close to a Maxwellian and, as a result, the non-thermal high energy tail is absent from the Comptonisation spectrum. 
Due to the effects of the pair production thermostat (Svensson 1984), the temperature of the electron plasma  is limited to  a few hundred keV. The temperature of the hot protons $kT_{\rm i}$ increases with $l_{\rm c}$ (from 0.04 to 9 MeV) in accordance with equation~(\ref{eq:temp}). 
\begin{figure*}
 \includegraphics[width=\textwidth]{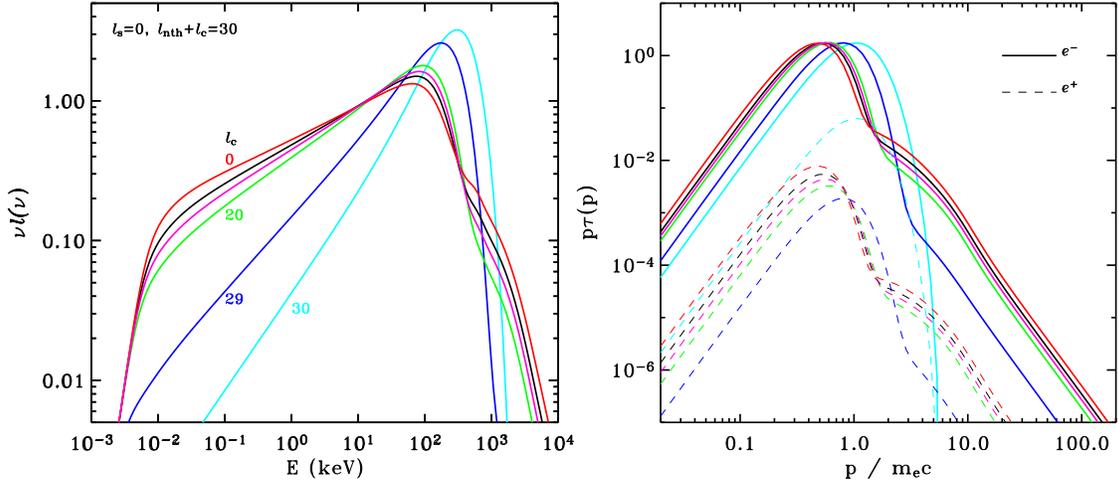}
 \caption{{ SSC models  with heating by hot protons.} Effects of varying the Coulomb heating compactness $l_{\rm c}$ at constant $l$ on the photon spectra  (left hand side) and the particle distributions (right hand side). The red curves shows our fiducial model obtained for $l_{\rm c}=0$, $l_{\rm nth}=30$, $l_B=10$, $\Gamma_{\rm inj}=3$, $\gamma_{\rm min}=1$, $\gamma_{\rm max}=1000$, $\tau_{\rm i}=2$, $R=5 \times 10^7$ cm. The other curves show a sequence of models obtained  keeping $l=l_{\rm c}+l_{\rm nth}=30$ while the Coulomb heating compactness $l_{\rm c}$  is varied .  The black, pink, green, blue and cyan curves show the results for $l_{\rm c}$=10, 15, 20, 29, 30 respectively. In the right hand side panel, the full curves show the electron distributions while the dashed ones show the positron distributions.
 }
 \label{fig:lcoul}
\end{figure*}

\section{Application to Cygnus X-1}\label{sec:cygx}
In the following we  present preliminary qualitative comparisons (these are not formal spectral fits)  of our simulation to the LHS and HSS broad-band high energy spectra of Cygnus X-1.
\subsection{Pure non-thermal acceleration models}

We find that a pure non-thermal injection model (without hot protons) provides a good description of the high energy spectra of Cygnus X-1 in both spectral states (see Fig~\ref{fig:lcygdat}).  According to our 'best fit'  models (see Table~\ref{tab:cyg}), the non-thermal compactness of the corona is comparable ($l_{\rm nth}\simeq5$)  in both spectral states. As expected most of the differences between hard and soft state are due to a change in the soft photon compactnes $l_{\rm s}$ that we assumed to be 0 and about 3$l_{\rm nth}$ in the soft state.
However the other parameters also have to change in order to fit the details of the spectrum. As already mentioned in section~\ref{sec:intro}, the Thomson optical depth $\tau_{\rm T}$ is larger in the hard state.
In the HSS the slope of the injected electrons is strongly constrained by the shape of the hard X-ray photon spectrum and must rather flat ($\Gamma_{\rm inj}\simeq2.1$). 
In the LHS on the contrary, the slope of the injected electrons must be steeper, otherwise the soft cooling photon flux is to strong and it is not possible to produce the observed high LHS temperature.   

For the same reason, the magnetic field must be lower in the hard state (while it is not very well constrained in the HSS but could be high). In fact if we compare the magnetic to radiation energy density in both models we find the  $U_{B}/U_{\rm R}\simeq3$ in the HSS while in the LHS,  $U_{B}/U_{\rm R}\simeq0.3$.

In any case, the magnetic field infered from our model in the LHS is probably an upper limit on the actual magnetic field in the source. The presence of external soft photons would imply a lower $B$ as mentioned just above. If Coulomb heating is important then $l_{\rm nth}$ must be lower to conserve  the same luminosity. At the same time, in order to keep the same non-thermal Compton luminosity it would be necessary to reduce the synchrotron losses and probably decrease $B$.
The fact that this maximum magnetic field  appears significantly below equipartition with radiation suggests that the emission of the corona is not powered by the magnetic field, as assumed in most accretion disc corona models (see e.g.  di Matteo, Celotti \&  Fabian 1997; Merloni \& Fabian 2001).

{ We stress however that this
requirement of a low magnetic field in the LHS is entirely due to  the observed strength of the non-thermal  MeV excess in the spectrum of Cygnus X-1  and our interpretation of it as the non-thermal tail of the Comptonising electron distribution.  If contrary to our one-zone hypothesis, the  MeV excess is produced in a different region than the bulk of the thermal Compton emission, the constraint might be  relaxed.}

\subsection{Models with Coulomb heating}
\begin{table*}
 \centering
  \caption{Models parameters of the simulations providing a good representation of the Cygnus X-1 spectra in the LHS and HSS (as indicated in the first row). The two first models are the pure non-thermal acceleration models shown in Fig. 4. The two others include heating by hot protons.  In addition to the model parameters the resulting proton temperature as well as the temperature of the thermal component of the electron distribution obtained from a fit with a Maxwellian are given.}
  \begin{tabular}{cccccccccccc}
  \hline
    state         & $l_{\rm nth}$  & $l_{\rm c}$ & $l_{\rm s}$ & $l_{B}$ & $\tau_{\rm i}$  & $\Gamma_{\rm inj}$ & $\gamma_{\rm min}$   &$\tau_{\rm T}$ &  $kT_{\rm e}$ (keV) & $kT_{\rm i}$ (MeV) \\
\hline
LHS    &     4.75     &  0           &    0          &   0.475   &   1.45    &         3.50               &             1          &    1.45    &    76        &         \\%060
 HSS    &   5.50     &   0          &    16.7     &      17.6 &    0.11      &        2.10               &            1           &    0.12     &    48                     &      \\%049
 \hline
 LHS   &    1.00     &   3.00     &    0.29    &    0    & 1.45            &         3.00                &          1.5                                      &   1.45    &     85      &   1.0 \\%041

HSS     &  1.98      & 0.96      &    17.6     &      0      &    0.11       &        2.64               &          1.5          &    0.12     &     67                     &     50 \\%198
 \hline
\end{tabular}
\label{tab:cyg}
\end{table*}
Alternatively, models whith heating by hot protons also provide a very good description of the spectra of Cygnus X-1. However even in these models some level of non-thermal acceleration is required in order to reproduce the non-thermal MeV tails. In our 'best fit' models about 25 \% of the heating power is provided in the form of non-thermal acceleration in the LHS while this fraction rises to about 2/3 in the  HSS. $l_{\rm nth}$ is larger by a factor of 2  in the soft state. For simplicity in this comparison we neglected the effects of magnetic field (and set $l_{B}=0$). The Coulomb compactness $l_{\rm c}$  is lower by a factor of three in the HSS but  since the optical depth is also lower this translates into a higher proton temperature  (about 50 MeV in the HSS vs only 1.3 MeV in  the LHS). In terms of proton to electron   temperature ratio $T_{\rm i}/T_{\rm e}\simeq10^3$ in the HSS and $T_{\rm i}/T_{\rm e}\simeq10$.

Therefore we find that in the LHS the proton temperature is significantly lower than in the standard 2-temperature accretion flow solutions (the temperature of the hot protons in typical ADAF models is in the range 10--100 MeV). In fact, this result can be obtained analytically. Indeed, it is based only on the properties of Coulomb coupling between electrons and protons. In the case of a pure Maxwellian plasma, the dependence of the Coulomb compactness on electron and proton temperatures can be approximated as follows (Dermer 1986): 
\begin{equation}
l_{\rm c}=\sqrt{8\pi}\tau_{\rm T}\tau_{\rm i}\ln\Lambda(\theta_{\rm i}-\frac{m_{\rm e}}{m_{\rm i}}\theta_{\rm e})(\theta_{\rm e}+\theta_{\rm i})^{-3/2},
\label{eq:coul}
\end{equation}
where $\ln\Lambda\simeq20$ is the Coulomb logarithm and $\theta_{\rm e}=\frac{kT_{\rm e}}{m_{\rm e}c^2}$ is the electron temperature in units of the electron rest mass. Equation~\ref{eq:coul} is valid in the limit of small $\theta_{\rm i}$ and $\theta_{\rm e}$.  To the first order in $\theta_{\rm i}/\theta_{\rm e}$ equation~(\ref{eq:coul}) can be rewritten as:
\begin{equation}
\frac{T_{\rm i}}{T_{\rm e} }\simeq1+\frac{m_{\rm i}}{m_{\rm e}}\frac{g l_{\rm c} }{\sqrt{56\pi}\ln\Lambda\tau_{\rm T}\tau_{\rm i}}\simeq 1+7\frac{gl_{\rm c}}{\tau_{\rm T}\tau_{\rm i}},
\label{eq:temp}
\end{equation}
where $g(\theta_{\rm e})=\sqrt{7\theta_{\rm e}}$ ($\simeq1$ in the LHS). The proton temperatures provided by equation~(\ref{eq:temp}) agree with the values obtained from our numerical simulations  within 30 percent, even when the lepton distribution strongly departs from a Maxwellian. 
For the observed luminosity and optical depth in the hard states of Cyg X-1, GX339-4, XTE J1118,
having $T_{\rm i}/T_{\rm e}>100$ as in typical ADAF solutions would requires an extremely compact emitting region of radius $R$ lower than a few gravitational radii. The main reason for this result is that the observed optical depth in these sources is larger than unity and therefore coulomb collisions are efficient at transferring energy from the ions to the  electrons.

\begin{figure}
\centering
 \includegraphics[width=8cm]{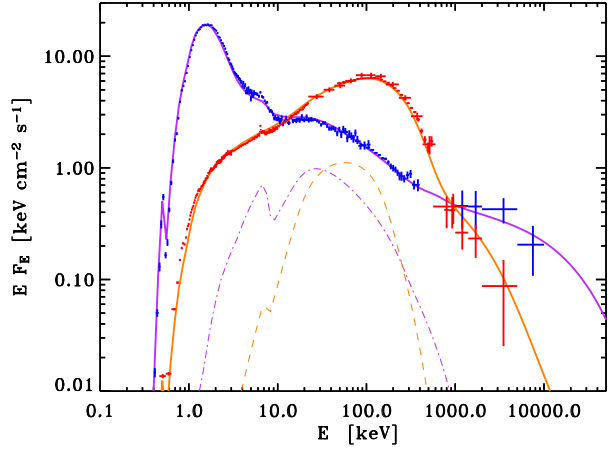}
 \caption{A comparison of the average \gro\ spectra for the HSS (blue)  and  LHS (red) of Cygnus X-1 from MC02, with models involving only injection of non-thermal particles as sole heating mechanism. At low energy the \gro\ data are complemented by \sax\ data (see MC02). 
 The thick purple line shows our absorbed soft state model which parameters are given in Table~1 while the orange line shows the hard state model of the same table. Reflection components were added to both spectra and are shown by the thin dot-dashed and dashed curves for the soft state model and the hard state model respectively.\label{fig:lcygdat}}
\end{figure}

\section{Conclusion}

We have explored the effects of the synchrotron self-absorption on the particle energy distribution and emitted radiation of the corona of accreting black holes. We have found that, in both spectral states of black hole binaries the 
 coronal emission  could be powered by a similar non-thermal acceleration mechanism. In the LHS the synchrotron and $e$-$e$ Coulomb boilers redistribute the energy of the non-thermal particles to form and keep a quasi-thermal electron distribution at a relatively high temperature, so that most of the luminosity is released through quasi thermal comptonisation. In the HSS,  the soft photon flux from the accretion disc becomes very strong and cools down the electrons, reducing the thermal Compton emissivity. This change in the soft photon flux could be caused either because the inner radius of the truncated disc moves inward into the central a hot accretion flow, or, in the framework of accretion disc corona model, because the disc temperature increases dramatically. Then most of the emission is produced by disc photons up-scattered by the non-thermal cooling electrons. Another difference between the two states is that the slope of the injected electrons has to be steeper in the LHS to reduce the synchrotron emission.   
 
Our comparison of simulations with the high energy spectra of Cygnus X-1 in the LHS allowed us to set upper limits on the magnetic field and the proton temperature.  Our results indicates that the  magnetic field is below equipatition with radiation (unlike what is assumed in most accretion disc corona models). The proton temperature is found to be significantly lower than predicted by standard 2-Temperature accretion flow models ($kT_{\rm i}\lesssim$2 MeV). 
{The constraint on the proton temperature is independent of the synchrotron boiler effects (it is valid even if the magnetic field B=0)}.  Indeed, in a two-temperature plasma, due to electron ion coupling, there is a relationship between the luminosity, the size of the emitting region,  the ion and electron temperatures, and the Thomson optical depth. For the measured luminosity, electron temperature and optical depth of several black hole binaries, like XTE~J1118+480 or GX~339--4, a proton temperature much higher than the electron temperature $T_{\rm i}/T_{\rm e}\sim$100--1000 (as in a two-temperature accretion flow) requires an extremely small size of the emitting region (a few $R_{\rm G} $ or less).
In the HSS the constraints on the magnetic field and the proton temperature are less stringent. The magnetic field could  power the emission of the corona.

\end{document}